\documentclass{article}
\usepackage{amsmath}
\usepackage{graphicx, amsfonts, authblk}
\usepackage[a4paper, margin=1.2in]{geometry}
\title{Auto-Balancer: Harnessing idle network resources for enhanced market stability}
\author{Arman Abgaryan, Utkarsh Sharma\footnote{\textbf{Acknowledgment}: The authors thank Joshua Tobkin for his invaluable guidance and support in this work.}}\affil{Supra DeFi Research}
\date{January 2025}

\begin{document}

\maketitle

\begin{abstract}
    We propose a mechanism embedded into the foundational infrastructure of a blockchain network, designed to improve the utility of idle network resources, whilst enhancing market microstructure efficiency during block production by leveraging both network-owned and external capital. By systematically seeking to use idle network resources for internally capture arbitrageable inefficiencies, the mechanism mitigates extractable value leakage, reduces execution frictions, and improves price formation across venues. This framework optimises resource allocation by incentivising an ordered set of transactions to be identified and automatically executed at the end of each block, redirecting any realised arbitrage income - to marketplaces operating on the host blockchain network (and other stakeholders), which may have otherwise been extracted as rent by external actors. Crucially, this process operates without introducing additional inventory risk, ensuring that the network remains a neutral facilitator of price discovery. While the systematic framework governing the distribution of these internally captured returns is beyond the scope of this work, reinvesting them to support the ecosystem deployed on the host blockchain network is envisioned to endogenously enhance liquidity, strengthen transactional efficiency, and promote the organic adoption of the blockchain for end users. This mechanism is specifically designed to operate on Supra's blockchain, leveraging its automation framework to enhance the network's efficiency.
\end{abstract}

\section{Introduction}
    The efficiency of markets operating on the host blockchain network depends on their ability to facilitate seamless transactional execution, thereby impacting price discovery, liquidity allocation, and overall market microstructure. However, during routine operations, user transactions in the markets can inadvertently create arbitrage opportunities. While external agents which contribute to reasserting market equilibrium \cite{hansson2022arbitrage} by correcting inadvertent price deviations provide valuable contribution to the market microstructure, their activities may cost marketplaces operating on the host blockchain network (for e.g., through adverse selection and impermanent loss \cite{aigner2021uniswap}). Further, the reliance on external arbitrageurs may introduce latency in execution, exacerbates price impact, and increases the risk of network congestion due to competitive transaction bidding. From the perspective of the host blockchain network, suboptimal utilisation of computational resources presents another inefficiency: while congestion can degrade user experience and increase transaction costs, under-utilisation represents an opportunity cost in terms of foregone revenue and economic inefficiency.\\
    \\
    We propose a framework that jointly addresses these inefficiencies by enabling the host blockchain network to discover a set of ordered transactions which would internalise arbitrage opportunities emerging during block production, while simultaneously leveraging under-utilised computational resources. By executing these balancer transactions, which are initiated and executed by the host blockchain network rather than external entities, the proposed mechanism captures and redistributes profits that would otherwise be taken by external arbitrageurs. This improves capital efficiency and strengthens market microstructure. Simultaneously, the framework decreases reliance on independent arbitrage bots, potentially reducing execution latency while ensuring that arbitrage-derived profits are equitably distributed among network participants. By capturing arbitrageable value, a portion of gains can be directed toward marketplaces operating on the host blockchain network, mitigating impermanent loss and improving market-making incentives. Furthermore, by seeking to synchronise on-chain prices across trading venues, the protocol minimises cross-market fragmentation, reduces transient pricing inefficiencies, and improves execution quality for traders and liquidity providers alike. These enhancements not only improve economic sustainability and systemic efficiency but also enhance the profitability of price-sensitive protocols operating on the network, while preserving a seamless and predictable trading environment.\\
    \\
    Embedding the automated arbitrage monetisation process directly within the network’s block production mechanism enhances market stability and improves the resilience of price-sensitive applications. By systematically responding to mispricing conditions in real time, the protocol fosters a more liquid, stable, and efficient trading environment, while simultaneously diminishing the need for discretionary intervention from external agents. This automated approach to market efficiency reduces information asymmetry, and reduces the risk of transaction ordering manipulation. Furthermore, by directing arbitrage-derived revenues to the protocol’s treasury—and redistributing them to price-sensitive applications operating on the host blockchain—the framework supports long-term ecosystem development, enhances security incentives, and fosters a more competitive, self-sustaining market infrastructure. 

\section{Objectives}
    The proposed protocol aims to enhance the efficiency of markets operating on the host blockchain network, by internalising arbitrage opportunities within block production. In doing so, it pursues two primary objectives: (i) minimising emerging price discrepancies to improve market efficiency; and (ii) optimising network resource utilisation.\\
    \\    
    For each market $i \in \mathbf{M}$, let $P^a_i(\mathbf{x}) \in \mathbb{R}^{+}$ be the effective price (of an asset $a$) after executing a set of transactions $\mathbf{x}$ in the network,  the arbitrageable price discrepancy ($\Delta_{ij}^a(\mathbf{x}) \in \mathbb{R}^{+}$) between markets $i$ and $j$ is defined as:
    \begin{equation*}
        \Delta_{ij}^a(\mathbf{x}) = \left| P^a_i(\mathbf{x}) - P^a_j(\mathbf{x}) \right|.
    \end{equation*}

    \noindent
    We seek to minimise the cumulative price discrepancy:
    \begin{equation*}
        \min_{\mathbf{x} \in \mathcal{X}} \quad \mathbb{E}\left[\sum_{i,j \in \mathbf{M}, i \neq j} \sum_{a \in \mathbf{A}} \Delta^a_{ij}(\mathbf{x})\right],     
    \end{equation*}

    \noindent
    where $\mathcal{X}$ denotes the set of feasible transaction sequences that satisfy condition $\mathbf{K}$:
    \begin{equation*}
        \mathcal{X} = \{ (\mathbf{x}_1, \mathbf{x}_2, \dots, \mathbf{x}_n) \mid \mathbf{K}(\mathbf{x}_1, \mathbf{x}_2, \dots, \mathbf{x}_n) = 1\}
    \end{equation*}
    \noindent
    where $\mathbf{K}$ is a binary condition function which returns 1 if the sequence is feasible based on technical convenience and economic utility of the submitted transactions; and  $\mathbf{A}$ denotes the universe of assets.\\
    \\
    Separately, and not necessarily combined with the preceding objective, we also seek to optimise the utilisation of the host blockchain network's resources. As such, let $w$ denote the amount of work ($w \in [0, C], \quad C \in \mathbb{R}^{+}$) - for e.g., computational, allocated within a block, and $C$ represent the block's maximum capacity, to define the block utilisation as:
    \begin{equation*}
        U(w) = \frac{w}{C}.
    \end{equation*}

    \noindent
    Now, to account for the marginal impact of additional work allocation to the network's performance, let $\psi(w)$ represent the performance cost incurred when executing $w$ work in a block. As such, to ensure that network performance remains acceptable, we impose a probabilistic constraint ($\delta$):
    \begin{equation*}
        \mathbb{E}[\psi(w)] \leq \delta.
    \end{equation*}

    \noindent
    This leads us to assemble the resource optimisation problem as follows:
    \begin{equation*}
    \begin{aligned}
        \max_{w \in [0, C]} \quad & U(w) = \frac{w}{C} \\
        \text{subject to} \quad & w \leq C, \\
        & \mathbb{E}[\psi(w)] \leq \delta.
    \end{aligned}
    \end{equation*}

    \noindent
    Now, since both objectives must be achieved simultaneously, the protocol addresses the following multi-objective optimisation problem:

    \begin{equation*}
        \min_{\mathbf{x} \in \mathcal{X}} \quad \lambda_1 \, \mathbb{E}\left[\sum_{i,j \in \mathbf{M}, i \neq j} \sum_{a \in \mathbf{A}} \Delta^a_{ij}(\mathbf{x})\right] - \lambda_2 \frac{\mathbf{w}(\mathbf{x})}{C},
    \end{equation*}
    
    \begin{equation*}
        \text{subject to} \quad \mathbb{E}[\psi(\mathbf{w}(\mathbf{x}))] \leq \delta.
    \end{equation*}

    \noindent
    where $\lambda_1, \lambda_2$ are weights balancing two objectives; $\mathbf{w}(\mathbf{x})$ is the work function, describing the amount computational work allocated within a block when executing a given set of transactions $\mathbf{x}$.\\
    \\    
    In essence, Auto-Balancer seeks to ensure that arbitrage transactions are optimally executed to reduce inter-market price discrepancies, while efficiently using the host blockchain network's idle computational resources.

\section{Algorithmic Framework}
    Auto-balancer is an endogenous framework integrated directly into the host blockchain network's (Supra) block production process. Its purpose is to utilise the host blockchain network's idle resources to internalise arbitrage opportunities that would otherwise be exploited by external agents. By doing so, it improves market efficiency, reduces the probability of losses related to extractable value emerging, and redistributes the gains to network stakeholders — without introducing inventory risk.\\
    \\
    The Auto-balancer is designed to internalise arbitrage opportunities within the host blockchain’s block production cycle by introducing balancer transactions - which focus on racing to capture emerging arbitrageable value, to be executed after user-initiated transactions, within a block. However, since the exact gas consumption of user-initiated transactions within a given block is not deterministically known ex-ante, balancer transactions must be ordered based on their expected contribution to arbitrage capture, as they are only executed using idle resources in the block. As such, transactions with a higher probability of capturing arbitrageable value should be assigned a greater likelihood of execution, ensuring efficient resource allocation within the block's computational constraints.\\
    \\
    We now explain the algorithmic framework in the following steps:
    
    \begin{itemize}
        \item (\textit{Step 1}) \textbf{User-initiated Submissions}: Users submit a set of transactions, 
        \begin{equation*}
            \mathcal{T}_u = \{T_{u,1}, T_{u,2}, \ldots, T_{u,M}\},
        \end{equation*}
        \noindent
        which are incorporated into the upcoming block. 
        
        \item (\textit{Step 2})\textbf{State Transition}: Once executed, these transactions transition the blockchain from state:
        \begin{equation*}
            \mathbf{S}_t = \{\mathbf{S}_t^{c_1}, \mathbf{S}_t^{c_2}, \ldots, \mathbf{S}_t^{c_m}\}
        \end{equation*}

        \noindent
        where $\mathbf{S}_t^{\mathbf{c}_1}$ is the set of states, associated with smart contract $c_1$ after $t$-th state update, which updates to:
        \begin{equation*}
            \mathbf{S}_{t+n} = \{\mathbf{S}_{t+n}^{c_1}, \mathbf{S}_{t+n}^{c_2}, \ldots, \mathbf{S}_{t+n}^{c_m}\},
        \end{equation*}

        \noindent
        capturing all changes in state variables resulting from user transactions.
        
        \item (\textit{Step 3}) \textbf{Balancer Transactions execution}: After user transactions, balancer transactions are executed, which query a reference market ($R$), for e.g. DFMM \cite{abgaryan2023dynamic}, for a price vector for $N$ assets:
        \begin{equation*}
            \mathbf{P}^R_{t+n} = \{p^R_{i,t+n}\}_{i=1}^{N},
        \end{equation*}
        \noindent
        and collecting prices from $J$ other venues:
        \begin{equation*}
            \mathbf{P}_{t+n} = \{\mathbf{P}^j_{t+n}\}_{j=1}^{J}, \quad \text{with} \quad \mathbf{P}^j_{t+n} = \{p^j_{i,t+n}\}_{i=1}^{N}.
        \end{equation*}

        \noindent
        Now, for each asset $i$ and venue $j$, the relative price deviation is computed as:
        \begin{equation*}
            \Delta p_{i,j,t+n} = \frac{p^j_{i,t+n} - p^R_{i,t+n}}{p^R_{i,t+n}}.
        \end{equation*}

        \noindent
        The balancer transactions execute arbitrage trades, if an arbitrage opportunity is detected, then - $|\Delta p_{i,j,{t+n}}| > \epsilon$, where $\epsilon \in \mathbb{R}^+$ is a threshold, using either flash loans\footnote{A flash loan is an uncollateralised, atomic loan that must be repaid within the same blockchain transaction. If repayment fails, the transaction reverts, ensuring no outstanding debt or risk to the lender.}, or network-owned liquidity.
        
        \item ({\textit{Step 4}}) \textbf{Transaction Set Discovery}: At the beginning of each execution window (an epoch), which is comprised of multiple blocks, an execution-priority ordered set of transactions $\mathbf{T} = \{T_1, T_2, \ldots, T_k\}$ is discovered, with the intention of seeking to restore any mispricing which may have emerged (between reference and other markets after execution of user initiated transactions) in the block-production process. This is achieved with the help of searchers, who essentially run a balancing function, of the form: 
         \begin{equation*}
            \mathbf{T}_{s,e} = \mathbf{O}_s(\mathbb{E}[\mathbf{S}_{e}], \mathbb{E}[\mathbf{r}_e], c),
        \end{equation*}

        \noindent
        where $\mathbf{S_e}$ is a matrix representing the set of cleared states after user transactions across blocks in $e$-th epoch; $\mathbf{r}_e$ is the set of available block capacity in each epoch; $c$ encapsulates conditions defined by governance (e.g., designated lending markets for flash loans or reference price sources); and $\mathbf{O}_s$ represents the searcher-specific function generating the optimal transaction set.\\
        \\
        At each epoch, the set of balancer transactions ($\mathbf{T}_e$) adjusts to market conditions and system states. The system incentivises searchers to propose optimal sets, which are evaluated via decentralised governance, to determine the applicable balancer transaction set in the next epoch. Selection criteria may include additional information submitted by searchers, for e.g.,
        \begin{enumerate}
            \item \textbf{Profit Estimate}: Expected arbitrageable value, i.e., $\mathbb{E}[\mathbf{\Pi}(\mathbf{T}_s)] \in \mathbb{R}^+$.
            \item \textbf{Gas Fee Estimate}: Expected gas cost for executing the arbitrage, i.e., $\mathbb{E}[\mathbf{G}(\mathbf{T}_s)] \in \mathbb{R}^+$.
            \item \textbf{Simulation}: An execution simulation using data from atleast the past few blocks, accounting for price impact, slippage, and historical inclusion rates (if available) to ensure economic viability and feasibility, under real network conditions.
            \item \textbf{Searcher Credibility}: Auto-balancer can assess searcher credibility over time via governance mechanisms to refine selection and decision-making.
        \end{enumerate}

        \noindent        
        In essence, searchers determine the order of transactions based on their ordering function, considering  an individual transactions $T_k$ financial output calculated as $\mathbb{E}[\mathbf{\Pi}(T_k)] - \mathbb{E}[\mathbf{G}(T_k)]$, where $\mathbb{E}[\mathbf{\Pi}(T_k)]$ is the expected arbitrage profit from transaction $T_k$; and $\mathbb{E}[\mathbf{G}(T_k)]$ is their expected gas cost for transaction $T_k$. And quite notably, the host blockchain network does not take any additional inventory risk in the entire process.
    \end{itemize}

\subsection{Reward Distribution}
    If $\mathbf{\Pi}(\mathbf{T}_e) \in \mathbb{R}^+$ represents the total arbitrage profit pool available for distribution among searchers ($\mathcal{S}$), marketplaces operating on the host blockchain network ($\mathcal{L}$), and the network's treasury ($\mathcal{N}$), then each group receives: $F_x = \omega_x \cdot \mathbf{\Pi}(\mathbf{T}_e), \quad x \in \{\mathcal{S}, \mathcal{L}, \mathcal{N}\}$, where the allocation weights $\omega_x$ are dynamically adjusted based on network conditions and satisfy: $\sum_{x \in \{\mathcal{S}, \mathcal{L}, \mathcal{N}\}} \omega_x = 1.$\\
    \\    
    The contribution of each marketplace $l \in \mathcal{L}$ to the arbitrage profit pool is captured by a function $\mathbf{\rho}(l)$, which quantifies the share of arbitrageable value facilitated by that marketplace. The marketplace-specific allocation can then be expressed as:
    \begin{equation*}
        F_l = \omega_{\mathcal{L}} \cdot \mathbf{\Pi}(\mathbf{T}_e) \cdot \frac{\mathbf{\rho}(l)}{\sum_{l' \in \mathcal{L}} \mathbf{\rho}(l')}, \quad l \in \mathcal{L},
    \end{equation*}

    \noindent
    ensuring that each marketplace's reward is proportional to its relative contribution to the overall arbitrage opportunity.\\
    \\
    For block producers, while a comprehensive incentive design for node operators is beyond the scope of this work, we will propose a mechanism to encourage fair execution of balancer transactions. Therein, block producers receive a fraction $\gamma \in (0,1)$ of the gas fees submitted with balancer transactions, ensuring they have a direct economic incentive to include them in blocks. On the other hand, to discourage manipulation, a slashing penalty is enforced if block producers fail to honour the prescribed execution order of balancer transactions.

\bibliography{main.bib}
\bibliographystyle{plain}

\end{document}